\documentclass[aps,prd,superscriptaddress,showpacs,amsmath,amssymb]{revtex4}
\usepackage{epsfig}
\usepackage{color}
\usepackage{latexsym}
\usepackage{amsmath}
\usepackage{amsfonts}
\usepackage{amssymb}
\usepackage{graphicx}
\usepackage[utf8]{inputenc}
\usepackage[T1]{fontenc}

\usepackage{graphicx}
\usepackage{amsfonts}
\newcommand{\be}{\begin{eqnarray}}
\newcommand{\ee}{\end{eqnarray}}

\begin{document}

\title{Statistical Mechanics of Wormholes}
\author{Paul H. Cox}
\email{phcox@tamuk.edu}
\affiliation{Department of Physics and Geosciences, Texas A\&M University-Kingsville, M.S.C. 175, Kingsville, TX 78363-8202, USA}
\author{Benjamin C. Harms}
\email{bharms@bama.ua.edu}
\affiliation{Department of Physics and Astronomy, The University
of Alabama, Box 870324, Tuscaloosa, AL 35487-0324, USA}
\author{Shaoqi Hou}
\email{shou@crimson.ua.edu}
\affiliation{Department of Physics and Astronomy, The University
of Alabama, Box 870324, Tuscaloosa, AL 35487-0324, USA}
\begin{abstract}
The statistical mechanics of a gas of Einstein-Kalb-Ramond wormholes is studied in this paper.  The wormholes studied are the result of sewing together two Reissner-Nordstrom-type black hole metrics at their horizons.  By requiring the stress-energy tensor associated with this geometry to be that of a Kalb-Ramond field, we obtain the mass and Kalb-Ramond `charge` of the wormholes in terms of the parameters describing the mass density, tension and pressure.  We investigate the statistical mechanics of this system of wormholes within the context of the statistical bootstrap model.  A gas of such wormholes is found to obey the bootstrap condition only for an extreme, non-thermodynamic, energy and `charge` distribution among the particles. We comment briefly on the scattering of quantum wormholes.
\end{abstract}
\pacs{04.20.Jb, 04.40.Nr,04.60.Bc}
\maketitle
%
\section{Introduction}
The concept of wormholes as classical or semiclassical models of particles is an old one.
Wormhole solutions of the Einstein equations were first studied by Einstein and Rosen \cite{ein} who called such solutions `bridges'.  The solutions in \cite{ein} were offered as models for the elementary particles known at the time (the proton and the electron).  This attempt and subsequent attempts to identify elementary particles as wormholes have failed.  Although no viable models of the known elementary particles as wormholes have been found, such  models may yet be found.  The possibility that wormholes may be created as a new form of matter also exists.  In \cite{cox} Kalb-Ramond (K-R) charged wormholes were shown to be quasi-stable against tunnelling  of gravitational instantons to the vacuum and to be potentially of relatively small mass in four space-time dimensions.  The wormholes studied in \cite{cox} can be constructed by sewing together two `charged' black hole metrics at their horizons.  The resulting space-time consists of two congruent regions which are geometrically connected but are causally disconnected. The stability of these wormholes was analyzed by calculating the Euclidian action obtained for the metric which arises from the condition of equivalence we place on the dynamical quantities associated with an Einstein-Kalb-Ramond geometry. In the present paper we treat a collection of such wormholes as an ideal gas and calculate the statistical mechanical quantities associated with this gas using a microcanonical ensemble appropriate to such a system.

In \cite{harms} the statistical mechanics of a gas of quantum black holes was developed. This development was shown in \cite{harms1} to be applicable to extended black objects such as strings and membranes in higher dimensional spaces.  This development was based on the statistical model of hadrons, and examined the conditions under which the bootstrap \cite{hage,frau,carl} relation holds,
\be 
\frac{\Omega(E)}{\rho(E)} \to 1\, , E \to \infty\, ,
\label{boot}
\ee
where $\Omega(E)$ is the microcanonical density of states and $\rho(E)$ is the degeneracy of states for an object which can exhibit excitations, e.g. a string.  In the context of the statistical model the  bootstrap constraint implied that hadronic resonances could themselves be considered to be composed of other resonances, thereby replacing the  elementary particle concept.	Scattering theory was later  developed and the scattering amplitudes were shown to possess the property of duality.  A by-product of the duality symmetry in the  scattering amplitudes is that the number of open channels in a  scattering process rises in parallel with the degeneracy of  states as the energy is increased \cite{frau}.    Black holes exhibit the string-like property (see for example \cite{bowick}) that their degeneracy of states increases exponentially, with the argument of the  exponential now quadratic in mass, at least to leading order in  a $U(1)$-charge expansion.    As an example, in natural units $(\hbar = c= G = 1)$, a  Schwarzschild (neutral) black hole has the following density  of states in mass space,  
\begin{eqnarray}  
\rho_{Schw} (m) \sim c\; e^{4\pi m^2}\, , 
\label{rhos} 
\end{eqnarray}  
where m is the mass of the black hole.	This result is a  non-perturbative quantum effect in the senses that it is obtained  from the WKB method and  that the argument of the exponential is of  order $\hbar^{-1}$(for a detailed derivation see  Ref.\cite{cole}).   
In \cite{cox} an expression similar to Eq.(\ref{rhos}) was obtained for E-K-R wormholes. We regard our expression for $\rho(E)$ as  the quantum degeneracy of wormhole states. In this  way the laws of quantum mechanics remain untouched and the  process of wormhole decay can be understood from an  S-matrix theory point of view.    The problem of  understanding the statistical mechanics of a gas of such  objects is a natural outcome of our identification of $\rho(E)$ as the quantum degeneracy of wormhole states.    In the following sections we shall find that the microcanonical  ensemble is the proper framework for analyzing this  problem.	The E-K-R wormholes studied in this paper satisfy the bootstrap constraint	(Eq.(\ref{boot})) in a sense described below, hence showing that a wormhole can be viewed as being made of other wormholes, very much as in the old statistical model  of hadrons.    The equilibrium state of such a system is not thermal.	A  quantum coherent view of wormhole decay (evaporation) can also  be obtained in a way analogous to strings.	Wormhole states  decay into other wormhole  states.	Although the problem of  constructing wormhole scattering amplitudes is beyond the scope of this paper, we shall nevertheless demonstrate that the number  of open channels for $n$-body decay of an E-K-R wormhole  does indeed grow precisely in parallel with the density of states, thereby allowing duality symmetry for wormhole  scattering amplitudes.	Wormholes may belong to a certain  class of string theories.    Obvious applications of the considerations presented in this  paper lie in cosmology and the very early universe, as well as  in galaxy formation.  
\label{intro}
\section{Charged Wormholes}
\label{charge}
\subsection{Metric Tensor Elements}
The invariant line element for each wormhole in a gas of wormholes is taken to be of the form 
\be 
ds^2 = -(1-\frac{b(r)}{r})\,dt^2 + \frac{1}{(1-\frac{b(r)}{r})}\,dr^2 + r^2\,d\theta^2 +r^2\,\sin^2(\theta)\,d\phi^2\, .
\label{metric}
\ee
In a local orthonormal coordinate basis, the proper reference frame of observers at fixed $r, \theta, \phi$, the basis vectors are
\be 
 e_{\hat{t}} &=& \left(1-b(r)/r\right)^{-1/2}\, e_t\: , \:\: e_{\hat{r}} = \left(1-b(r)/r\right)^{1/2}\, e_r\nonumber \\
\qquad
 e_{\hat{\theta}} &=&  e_{\theta}\: , \hspace{7mm} e_{\hat{\phi}} =  e_{\phi} \, .
\ee
In such a basis, pressure $p$ is given by the equal $\theta\theta$ and $\phi\phi$ components of the stress-energy tensor, while energy density $\rho$ and tension $\tau$ are given by the negatives of the $\hat{t}\hat{t}$ and $\hat{r}\hat{r}$ components respectively.    We assume the stress-energy tensor is that of a K-R field; for a static spherical-symmetric solution the K-R field will have only one independent non-zero component which then gives $\rho c^2 = \tau = p$.  
The requirement of equality of the energy density, tension, and pressure,  which are related to the stress-energy tensor components by
$T_{\hat{t}\hat{t}} = -\rho(r)\, c^2,\; T_{\hat{r}\hat{r}} = -\tau(r)$, and $T_{\hat{\theta}\hat{\theta}} = T_{\hat{\phi}\hat{\phi}} = p(r)$, leads to 
\be 
b(r) = b_0(1+A)- \frac{b_0^2\, A}{r} \, ,
\label{b}
\ee
where $b_0$ and $A$ are constants which determine the wormhole radius and the magnitude of the wormhole's apparent charge, respectively.  $A$ is limited to the range $0\leq\, A \,\leq 1$.  Substituting this form into Eq.(\ref{metric}) gives a Reissner-Nordstrom-type metric in which $b_0\,(1+A)$ is proportional to the wormhole mass $M$, and $b_0^2\,A$ is proportional to the square of the K-R `charge', $Q$.  The energy density, tension, and pressure are related to $b(r)$ by
\be 
\rho(r)\,c^2 = \tau(r) = p(r) = \frac{b(r)^{'}}{r^2} = \frac{b_0^2\, A}{r^4}\, ,
\label{rho}
\ee
where the prime indicates differentiation with respect to $r$.  This metric describes two Reissner-Nordstrom-type black holes glued together at their horizons (Fig.1) to form a wormhole of radius $b_0$ with a causal boundary at
\be 
r_+ = M + \sqrt{M^2-Q^2} = b_0\, .
\label{rp}
\ee
\begin{figure}[ht]
\centering
\includegraphics[viewport=0cm 0cm 25.59cm 20.34cm,scale=0.55,clip]{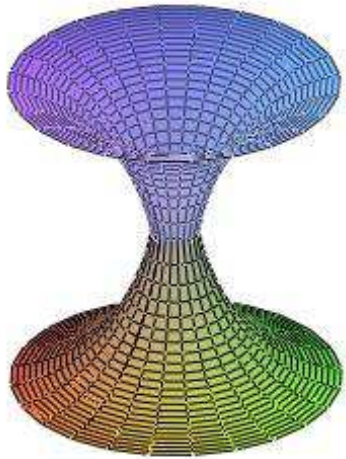}

\caption{Wormhole obtained by sewing together two Reissner-Nordstrom-type wormholes at their horizons}
\label{wh}
\end{figure}
\subsection{Einstein-Kalb-Ramond Action}

The metric in Eq.(\ref{metric}) is part of a solution for the Einstein-Kalb-Ramond action
\be 
S = \frac{1}{8\,\pi\,G}\,\int_M\sqrt{-g}\left(R -\frac{1}{4}\,H_{\mu\nu\lambda}\,H^{\mu\nu\lambda}\right)\,d^4x \, ,
\label{action}
\ee
where $g$ is the negative of the metric tensor determinant, $R$ is the Ricci scalar, and $H_{\mu\nu\lambda}$ is the totally antisymmetric K-R field.  To obtain the metric in Eq.(\ref{metric}), the K-R field is taken to have only one non-zero element, $H_{023}$, and
\be 
H_{\mu\nu\lambda}\,H^{\mu\nu\lambda} = -f(r)^2 = -\frac{b_0^2\,A}{r^4}\, .
\label{H}
\ee
$H_{\mu\nu\lambda}$ satisfies the field equation
\be 
\frac{1}{\sqrt{-g}}\,\partial_\mu\left(\sqrt{-g}\,H^{\mu\nu\lambda}\right) = 0\, .
\ee

To analyze the stability of wormholes described by the metric in Eq.(\ref{metric}) the action is calculated using the procedure of \cite{gib}.  Since the Ricci scalar is not quadratic in first derivatives, the calculation requires a boundary term which contributes to the action.  This term can be written as
\be 
S_B = \frac{1}{8\,\pi\,G}\int_{\partial\,M}\,\sqrt{-h}\,K\,d^3x\, ,
\label{bdA}
\ee
where $h$ is the induced metric on the boundary and $K$ is the trace of the second fundamental form (expressed using the boundary's unit normal $n_{\mu}$)
\be 
K_{\mu\nu} = \nabla_{\mu}n_\nu -n_\lambda\,n^\lambda\,n_\mu\,n^\sigma\,\nabla_\nu\,n_\sigma \, .
\label{sec}
\ee
The value of the action in Eq. (\ref{bdA}) must be renormalized to counter the divergence which arises when $r\,\to\,\infty$ (see \cite{cox} for more details).  The renormalized action is
\be 
S_B = \frac{1}{8\,\pi\,G}\int_{\partial\,M}\,\sqrt{-h}\,[K]\,d^3x\, ,
\label{bdAn}
\ee
where $[K]= K-K^F|_{r\to\infty}$ and $K^F$ is the second fundamental form for flat space.  In \cite{cox}
the total Euclidean action was shown to be (using $c\,=\,1$ units)
\be 
S_E = \frac{i\pi\,b_0^2\,(1+A/2)}{G\,(1-A)}\, .
\label{se}
\ee
\subsection{Quantum Density of States}
The degeneracy of states for a quantum wormhole can be found from the relation
\be 
\rho = c\,{\rm e}^{-i\,S_E/\hbar}\, ,
\label{rho}
\ee
where the coefficient $c$ is s constant which arises due to undetermined effects in the purely perturbative quantum field-theoretical sector of the theory.
Using the expression for $S_E$ in Eq.(\ref{se}), the expression for an E-K-R wormhole density of states has the form

\begin{equation}\label{rhow}
\rho=c\,\exp\Big[\frac{\pi b_0^2(1+A/2)}{G(1-A)}\Big]=c\,\exp\Big[\frac{\pi E^2(1+A/2)}{G(1-A)(1+A)^2}\Big]\, ,
\end{equation}
where $E=b_0(1+A)$, and the K-R `charge' is $Q=b_0\sqrt{A}$.  The fact that $\rho$ in Eq.(\ref{rhow}) increases with the exponential of $E^2$ shows that a gas of such wormholes cannot be described by a canonical ensemble.  A partition function cannot be defined for this system, because $\rho$ grows faster than the probability factor which decays only linearly with $E$.  Therefore the gas of wormholes must be described by means of a microcanonical ensemble.  

\subsection{Microcanonical Density of States}

The microcanonical density of states can be found from the relation
\be 
\Omega(E,Q,V) = \sum_n\,\Omega_n(E,Q,V) \, ,
\label{Om}
\ee
where $\Omega_n(E,Q,V)$ is the contribution of a system of $n$ wormholes to the total microcanonical density of states.  Because the two exterior regions of our wormhole solution are causally separated by a horizon, they are effectively separate particles as far as their relationship to neighboring particles is concerned.  In one respect the two regions are not identical; in one the K-R field $H_{\mu\nu\lambda}$ is dual to an inward-pointing vector; in the other exterior, the corresponding vector must point outward, thus defining 'negative' and 'positive' 'charges', respectively.  With this in mind, using the statistical model of a noninteracting gas, $n$ wormholes (actually here we count wormhole exteriors), $n_+$ positive and $n_-$ negative, yield a microcanonical density of states 
, $\Omega_n(E, Q, V)$, given by
\begin{eqnarray}
\Omega_n(E, Q, V)&=&\Big[\frac{V}{(2\pi)^3}\Big]^n\frac{1}{n!}\prod_{i_+=1}^{n_+}\prod_{i_-=1}^{n_-}\int dQ_{i_+}\int dE_{i_+}\rho(E_{i_+})\int dQ_{i_-}\int dE_{i_-}\rho(E_{i_-}) 
\nonumber\\
&&\delta\Bigg(\sum_{i_+=1}^{n_+}E_{i_+}+\sum_{i_-=1}^{n_-}E_{i_-}-E\Bigg)\delta\Bigg(\sum_{i_+=1}^{n_+}Q_{i_+}+\sum_{i_-=1}^{n_-}Q_{i_-}-Q\Bigg)\, ,
\label{On}
\end{eqnarray}
with $n_++n_-=n$. To evaluate the expression on the right hand side of Eq.(\ref{On}) the expression
\begin{equation}\label{eq:7}
\prod_{i_+=1}^{n_+}\prod_{i_-=1}^{n_-}\exp\frac{\pi}{G}\Big[\frac{b_{0i_+}^2(1+A_{i_+}/2)}{1-A_{i_+}}+\frac{b_{0i_-}^2(1+A_{i_-}/2)}{1-A_{i_-}}\Big]
\end{equation}
is evaluated by considering the function
\begin{equation}\label{eq:d}
f=\sum_{i_+=1}^{n_+}\frac{b_{0i_+}^2(1+A_{i_+}/2)}{1-A_{i_+}}+\sum_{i_-=1}^{n_-}\frac{b_{0i_-}^2(1+A_{i_-}/2)}{1-A_{i_-}}\, .
\end{equation}
When $f$ is maximum, the integrand in (\ref{On}) is also a maximum. The function $f$ must satisfy the two conditions which arise from the conservation of energy and charge  
\begin{eqnarray}\label{eq:10}
\left\{
\begin{array}{c}
\displaystyle\sum_{i_+=1}^{n_+}E_{i_+}+\displaystyle\sum_{i_-=1}^{n_-}E_{i_-}=\sum_{i_+=1}^{n_+}b_{0i_+}(1+A_{i_+})+\sum_{i_-=1}^{n_-}b_{0i_-}(1+A_{i_-})=E\\
\\
\displaystyle\sum_{i_+=1}^{n_+}b_{0i_+}\sqrt{A_{i_+}}-\displaystyle\sum_{i_-=1}^{n_-}b_{0i_-}\sqrt{A_{i_-}}=Q \, .
\end{array}
\right.
\end{eqnarray}

The method of Lagrangian multipliers can be used to find the extreme value of $f$. The constraints on $f$ are incorporated by defining the function
\be
L=f+\lambda_1\Big[\displaystyle\sum_{i_+=1}^{n_+}b_{0i_+}(1+A_{i_+})+\sum_{i_-=1}^{n_-}b_{0i_-}(1+A_{i_-})-E\Big]+\lambda_2\Big[\displaystyle\sum_{i_+=1}^{n_+} b_{0i_+}\sqrt{A_{i_+}}-\sum_{i_-=1}^{n_-} b_{0i_-}\sqrt{A_{i_-}}-Q\Big]\, . 
\label{L}
\ee
Extremizing $L$ with respect to the parameters gives
\begin{eqnarray}
&0=&\frac{\partial L}{\partial b_{0j_+}}=\frac{2b_{0j_+}(1+A_{j_+}/2)}{1-A_{j_+}}+\lambda_1(1+A_{j_+})+\lambda_2\sqrt{A_{j_+}}
\label{eq:1}\\
&0=&\frac{\partial L}{\partial A_{j_+}}=\frac{3b_{0j_+}^2}{2(1-A_{j_+})^2}+\lambda_1b_{0j_+}+\frac{\lambda_2b_{0j_+}}{2\sqrt{A_{j_+}}}
\label{eq:2}\\
&0=&\frac{\partial L}{\partial b_{0j_-}}=\frac{2b_{0j_-}(1+A_{j_-}/2)}{1-A_{j_-}}+\lambda_1(1+A_{j_-})-\lambda_2\sqrt{A_{j_-}}
\label{eq:3}\\
&0=&\frac{\partial L}{\partial A_{j_-}}=\frac{3b_{0j_-}^2}{2(1-A_{j_-})^2}+\lambda_1b_{0j_-}-\frac{\lambda_2b_{0j_-}}{2\sqrt{A_{j_-}}}
\label{eq:4}\\
&0=&\frac{\partial L}{\partial \lambda_1}=\sum_{i_+=1}^{n_+}b_{0i_+}(1+A_{i_+})+\sum_{i_-=1}^{n_-}b_{0i_-}(1+A_{i_-})-E\\
&0=&\frac{\partial L}{\partial \lambda_2}=\sum_{i_+=1}^{n_+} b_{0i_+}\sqrt{A_{i_+}}-\sum_{i_-=1}^{n_-} b_{0i_-}\sqrt{A_{i_-}}-Q
\end{eqnarray}

The multiplier $\lambda_1$ can be found in terms of the parameters for positively charged wormholes by taking the difference Eq.(\ref{eq:1})$\times(\sqrt{A_{j_+}})^{-1}-$Eq.(\ref{eq:2})$\times2\sqrt{A_{j_+}}/b_{0j_+}$ 
\begin{equation}\label{eq:b}
\lambda_1=b_{0j_+}\frac{A_{j_+}^2+4A_{j_+}-2}{(1-A_{j_+})^3}\, .
\end{equation}
Substituting $\lambda_1$ into Eq.(\ref{eq:1}) gives $\lambda_2$ 
\begin{equation}\label{eq:a}
\lambda_2=-b_{0j_+}\sqrt{A_{j_+}}\frac{2A_{j_+}^2+5A_{j_+}-1}{(1-A_{j_+})^3}\, .
\end{equation}
Dividing Eq.($\ref{eq:a}$) by Eq.(\ref{eq:b})gives the ratio of the Lagrange multipliers
\begin{equation}\label{eq:c}
\frac{\lambda_2}{\lambda_1}=-\sqrt{A_{j_+}}\frac{2A_{j_+}^2+5A_{j_+}-1}{A_{j_+}^2+4A_{j_+}-2}\, .
\end{equation}
Similarly for negatively charged wormholes, Eq.(\ref{eq:3})$\times(\sqrt{A_{j_-}})^{-1}$-Eq.(\ref{eq:4})$\times2\sqrt{A_{j_-}}/b_{0j_-}$ leads to
\begin{equation}\label{eq:6}
\lambda_1=b_{0j_-}\frac{A_{j_-}^2+4A_{j_-}-2}{(1-A_{j_-})^3}
\end{equation} 
for $\lambda_1$ and substituting $\lambda_1$ into Eq.(\ref{eq:3}), gives
\begin{equation}\label{eq:5}
\lambda_2=b_{0j_-}\sqrt{A_{j_-}}\frac{2A_{j_-}^2+5A_{j_-}-1}{(1-A_{j_-})^3}\, .
\end{equation}

The ratio of Eq.(\ref{eq:5}) to Eq.(\ref{eq:6}) is
\begin{equation}\label{eq:7}
\frac{\lambda_2}{\lambda_1}=\sqrt{A_{j_-}}\frac{2A_{j_-}^2+5A_{j_-}-1}{A_{j_-}^2+4A_{j_-}-2}
\end{equation}
The values of the Lagrange multipliers can be summarized as
\begin{eqnarray}
\left\{
\begin{array}{c}
\lambda_1=b_{0j_+}\displaystyle\frac{A_{j_+}^2+4A_{j_+}-2}{(1-A_{j_+})^3}=b_{0j_-}\frac{A_{j_-}^2+4A_{j_-}-2}{(1-A_{j_-})^3}\\
 \\
\lambda_2=-b_{0j_+}\sqrt{A_{j_+}}\displaystyle\frac{2A_{j_+}^2+5A_{j_+}-1}{(1-A_{j_+})^3}=b_{0j_-}\sqrt{A_{j_-}}\frac{2A_{j_-}^2+5A_{j_-}-1}{(1-A_{j_-})^3}\\
 \\
\displaystyle\frac{\lambda_2}{\lambda_1}=-\sqrt{A_{j_+}}\frac{2A_{j_+}^2+5A_{j_+}-1}{A_{j_+}^2+4A_{j_+}-2}=\sqrt{A_{j_-}}\frac{2A_{j_-}^2+5A_{j_-}-1}{A_{j_-}^2+4A_{j_-}-2}
\end{array}\, .
\right.
\end{eqnarray}

By analyzing the roots of the equations for $A_{j+}$ and $A_{j-}$ given in (\ref{eq:c}) and (\ref{eq:7}) the function $f$ can be shown to be unable to reach an extreme value in the range specified under the particular constraints Eq.(\ref{eq:10}).

In order to make $\Omega_n(E, Q, V)$ satisfy the bootstrap condition, one term in the function $f$ must be taken to be as large as possible and all others are as small as possible. When $Q\ge 0$, suppose the $n$th wormhole carries the energy $E_n=b_{0n}(1+A_n)$ and `charge' $Q_n= b_{0n}\sqrt{A_n}$, while the $n-1$ other wormholes carry energies $E_i=b_{0m}(1+A_m)$ and `charges' $Q_i=-b_{0i}\sqrt{A_i}$ ($i=1, 2, \cdots, n-1$), where $b_{0m}, A_m$ are the smallest values of the radius and the tension of a wormhole, respectively. The conservation relations for energy and `charge` are
\begin{eqnarray}\label{eq:k}
b_{0n}(1+A_n)+(n-1)b_{0m}(1+A_m)=E\\
b_{0n}\sqrt{A_n}-(n-1)b_{0m}\sqrt{A_m}=Q\label{eq:l}\, .
\end{eqnarray}
The microcanonical density of states is thus
\begin{eqnarray}\label{HOm}
\Omega_n(E, Q, V)=\Bigg[\frac{c\,V}{(2\pi)^3}\Bigg]^n\frac{1}{n!}\exp\Bigg[\frac{\pi}{G}\frac{b_{0n}^2(1+A_n/2)}{1-A_n}\Bigg]\exp\Bigg[(n-1)\frac{\pi}{G}\frac{b_{0m}^2(1+A_m/2)}{1-A_m}\Bigg]\, .
\end{eqnarray}
If $b_{0m}\approx 0, A_m\approx 0$, then $b_{0n}\approx b_0, A_n\approx A$, and $\Omega_n(E,Q,V)$ can be written approximately as
\begin{eqnarray}
\Omega_n(E, Q, V)\approx\Bigg[\frac{c\,V}{(2\pi)^3}\Bigg]^n\frac{1}{n!}\exp\Bigg[\frac{\pi}{G}\frac{b_{0}^2(1+A/2)}{1-A}\Bigg]\, .
\end{eqnarray}
 
The most probable equilibrium configuration is the one satisfying the condition,
\begin{equation}
\frac{d\Omega_n(E, Q, V)}{dn}\Bigg|_{n=N(E, Q, V)}=0 \, .
\end{equation}
Using Eq.(\ref{eq:k}) and Eq.(\ref{eq:l}),the value of $N(E,Q,V)$ can be calculated from the expression
\begin{eqnarray}\label{eq:m}
\exp[\Psi(N+1)]=\frac{c\,V}{(2\pi)^3}\exp\Bigg[\frac{\pi}{G}\frac{b_{0m}^2(1+A_m/2)}{1-A_m}-\frac{\pi}{G}F(b_{0N}, A_N)\Bigg]\, ,
\label{Psi}
\end{eqnarray}
with 
\begin{displaymath}
F(b_{0N}, A_N)=\frac{(2A_N^2+5A_N-1)\sqrt{A_NA_m}+(A_N^2+4A_N-2)(1+A_m)}{(1-A_N)^3}.
\end{displaymath}
$\Omega_N(E,Q,V)$ then satisfies the bootstrap condition, Eq. (\ref{boot}), if the value of $N$ is such that $\displaystyle\Bigg[\frac{c\,V}{(2\pi)^3}\Bigg]^N\frac{1}{\Gamma(N+1)}$ equals the constant $c$. 

\subsection{Microcanonical Thermodynamics}
The thermodynamic quantities associated with the microcanonical ensemble can be calculated from the expression in Eq.(\ref{Om}).  The total entropy is thus
\begin{eqnarray}
S(E, Q, V)&\equiv& \ln\Omega(E, Q, V)\approx\ln\Omega_N(E, Q, V)
\nonumber\\
&\approx&N\ln\Bigg[\frac{V}{(2\pi)^3}\Bigg]-\ln\Gamma(N+1)+\frac{\pi}{G}[f_N+(N-1)f_m]\, ,
\end{eqnarray}
where $f_i=\displaystyle\frac{b_{0i}^2(1+A_i/2)}{1-A_i}, i=N, m$. The inverse temperature $\beta$ is obtained from the total entropy according to 
\begin{eqnarray}
\beta&=&\frac{\partial S}{\partial E}
=\frac{\pi}{G}\frac{\partial f_N}{\partial E}
=\frac{\pi}{G}\frac{b_{0N}(-A_N^2-4A_N+2)}{(1-A_N)^3}\, .
\end{eqnarray}
The microcanonical specific heat is 
\begin{equation}
C_V=-\beta^2\frac{dE}{d\beta}=-\frac{\pi}{G}\frac{b_{0N}^2(A_N^2+4A_N-2)^2}{(1-A_N)(2-10A_N+23A_N^2+3A_N^3)}\, ,
\end{equation}
showing that the microcanonical specific heat is always negative, which is allowed for the microcanonical ensemble.
The pressure $P$ can be obtained by differentiating the entropy with respect to the volume, giving
\begin{equation}
\beta P=\frac{\partial S}{\partial V}=\frac{N}{V}\, ,
\end{equation}
which is expression for the pressure of an ideal gas. 

If $Q<0$, the signs on the individual charges are reversed, so that $Q_n=-b_{0n}\sqrt{A_n}$ and $Q_m=b_{0m}\sqrt {A_m}$. Eq.(\ref{eq:l}) then becomes
\begin{displaymath}
-b_{0n}\sqrt{A_n}+(n-1)b_{0m}\sqrt{A_m}=Q=-Q'\, ,
\end{displaymath}
which has a form similar to that of Eq.(\ref{eq:l}). 
\subsection{Scattering of Quantum Wormholes}
Wormholes, like black holes, have a degeneracy of states which grows as the exponential of the energy squared.  Such a dependence of the degeneracy of states on the energy is characteristic of geometrically extended objects such as strings and branes.  Such objects have scattering amplitudes which possess the property of duality.  As shown in \cite{frau}, the duality property implies that the number of $n$-body channels open in the statistical model used above must increase in parallel with the number of resonances as the center-of-mass energy is increased.   The number of open $n$-body channels is 
\be 
N_n \sim \rho(E)\, , \hspace*{3mm}E \to \infty
\label{nchan}
\ee
In the statistical model the number of resonances is
\be 
N_n = \frac{c^n}{n!}\,\prod_{i=1}^n\,\int_{E_0}^{E-E_{i+1}}\,dE_i\,\rho(E_i)\, ,
\label{oc}
\ee
where $E_{n+1} = E_0$.  The contributions from each integral will be small for all $E_i \approx E$ except for one, say $i = n$.  This is an indication that the one wormhole possesses most of the energy, and the remaining $n-1$ share whatever small amount is left.   Thus if $E>>E_0$, the right hand side of Eq.(\ref{oc}) will have the same form as Eq.(\ref{nchan}), showing that the wormhole scattering amplitudes possess the property of duality.
\section{Conclusions}
The use of the microcanonical ensemble to analyze the statistical mechanics of a gas of wormholes is necessary, because as shown in Section II.C the form of the quantum mechanical density of states precludes the formulation of a partition function.  The latter is a requirement for a system of particles in thermodynamic equilibrium.  As was the case for a gas of microscopic black holes \cite{harms}, a gas of wormholes cannot be in thermodynamic equilibrium.  Most of the energy of the gas resides in one of the wormholes, and the remaining constituents of the gas share the very small remainder of the total energy of the system. 
\par
In the limit that one wormhole in the gas carries nearly all of the energy and `charge', the gas obeys the bootstrap condition in Eq.(\ref{boot}).  The quantum corrections to the degeneracy of states $\rho(E)$, which are contained in the constant $c$ in Eq.(\ref{rho}), have a significant effect on the determination of the most probable equilibrium condition, as can be seen from Eq.(\ref{Psi}). 
\par
In \cite{cox} the possibility that the mass of the wormhole could be very small without violating the reality condition on $r_+$ was shown to hold for E-K-R wormholes. If, for example, the constituents of a gas of wormholes can be considered as a semiclassical description of K-R axions, the wormhole masses are in the range (depending on the interactions allowed with other particles) $10^{-2}$ eV to 25 MeV.  Such wormholes could be created by high energy cosmic rays or in an accelerator,  and they are quasi-stable, allowing the formation at least briefly of a wormhole gas.     
\acknowledgments
This work was supported in part by the U.S. Department
of Energy under Grant No. DE-FG02-10ER41714 .

\end{document}